\documentclass[10pt,a4paper,journal,twocolumn]{IEEEtran}
\usepackage{cite,graphicx,amsmath}
\usepackage{color}

\title{Multisource Self-calibration for Sensor Arrays}
\author{Stefan J. Wijnholds~\IEEEmembership{Student member,~IEEE} and Alle-Jan
  van der Veen~\IEEEmembership{Fellow,~IEEE}
\thanks{Copyright (c) 2008 IEEE. Personal use of this material is
  permitted. However, permission to use this material for any other purpose
  must be obtained from the IEEE by sending a request to
  pubs-permissions@ieee.org.}
\thanks{This work was supported by the Netherlands Institute for Radio
  Astronomy (ASTRON) and by NWO-STW under the VICI programme (DTC.5893).}
\thanks{S.J. Wijnholds is with ASTRON, Dwingeloo, The Netherlands. A.-J. van
  der Veen is with the Delft University of Technology, Delft, The
  Netherlands. Email: wijnholds@astron.nl, a.j.vanderveen@tudelft.nl}}

\def\revdate{manuscript, revised 5 February 2009}
\def\argmin{\mathrm{argmin}}
\def\diag{\mathrm{diag}}
\def\trace{\mathrm{tr}}
\def\vecdiag{\mathrm{vecdiag}}
\def\jcmplx{\mathrm{j}}
\def\e{\mathrm{e}}
\def\expect{\mathcal{E}}
\def\vec{\mathrm{vec}}

\def\bc{\mathbf{c}}
\def\bg{\mathbf{g}}
\def\bu{\mathbf{u}}
\def\by{\mathbf{y}}

\def\bA{\mathbf{A}}
\def\bC{\mathbf{C}}
\def\bG{\mathbf{G}}
\def\bI{\mathbf{I}}
\def\bJ{\mathbf{C}}
\def\bM{\mathbf{M}}
\def\bR{\mathbf{R}}
\def\bW{\mathbf{W}}
\def\bY{\mathbf{Y}}

\def\bgamma{{\mbox{\boldmath{$\gamma$}}}}
\def\btheta{{\mbox{\boldmath{$\theta$}}}}
\def\bsigma{{\mbox{\boldmath{$\sigma$}}}}
\def\bphi{{\mbox{\boldmath{$\phi$}}}}

\def\bGamma{{\mbox{\boldmath{$\Gamma$}}}}
\def\bLambda{{\mbox{\boldmath{$\Lambda$}}}}
\def\bSigma{{\mbox{\boldmath{$\Sigma$}}}}
\def\bPhi{{\mbox{\boldmath{$\Phi$}}}}

\begin{document}
\maketitle

\begin{abstract}
  Calibration of a sensor array is more involved if the antennas have
  direction dependent gains and multiple calibrator sources are simultaneously
  present. We study this case for a sensor array with arbitrary geometry but
  identical elements, i.e. elements with the same direction dependent gain
  pattern. A weighted alternating least squares (WALS) algorithm is derived
  that iteratively solves for the direction independent complex gains of the
  array elements, their noise powers and their gains in the direction of the
  calibrator sources. An extension of the problem is the case where the
  apparent calibrator source locations are unknown, e.g., due to refractive
  propagation paths. For this case, the WALS method is supplemented with
  weighted subspace fitting (WSF) direction finding techniques. Using Monte
  Carlo simulations we demonstrate that both methods are asymptotically
  statistically efficient and converge within two iterations even in cases of
  low SNR.
\end{abstract}


\IEEEpeerreviewmaketitle

\section{Introduction}

In this paper we study the calibration of the direction dependent response of
sensor array antennas, excited by simultaneously present calibrator
sources. The antenna array has arbitrary geometry but identical antennas.  The
calibration involves the complex gain of the antenna elements towards each
source.  The source powers are known but we will allow for small deviations in
apparent source locations to account for, e.g., refractive propagation
paths. This problem is one of the main challenges currently faced in the field
of radio astronomy.  For low frequency observations ($<$ 300 MHz) this
community is building or developing a number of new instruments, for example
the low frequency array (LOFAR) \cite{Bregman2005-1}, the Murchison wide field
array (MWA) \cite{Lonsdale2008-1} and the primeval structure telescope (PaST)
\cite{Peterson2005-1}, which are all large irregular phased arrays. These
difficulties arise due to the influence of the ionosphere on the propagation
of radio waves, which can qualitatively be categorized into the following four
regimes depending on the field of view (FOV) of the individual receptors and
the baselines between them \cite{Lonsdale2004-1}:
\begin{figure}
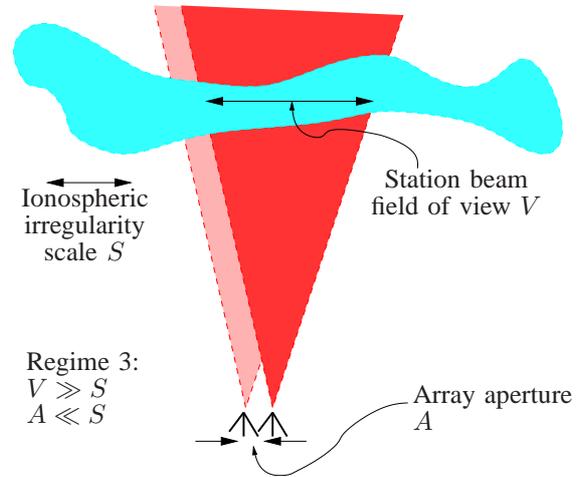

  \begin{center}
    \input fig1   
  \end{center}
  \caption{Graphical representation of regime 3: all lines of sight towards a
    single source sample the same ionosphere, but the ionospheric delay
    differs per source due to the large FOV and short baselines (after
    \cite{Lonsdale2004-1}).}
  \label{fig:lonsdale3}
\end{figure}
\begin{enumerate}
\item All antennas and all lines of sight sample the same ionospheric delays,
  thus the ionosphere causes no distortion of the array manifold (small FOV,
  short baselines).
\item Lines of sight from different antennas sample different ionospheric
  patches, but all sources in the antenna FOV experience the same delay; the
  ionosphere thus causes an antenna based gain effect (small FOV, long
  baselines).
\item All lines of sight towards a single source sample the same ionosphere
  but the delay differs per source. This regime requires source based
  direction dependent calibration (large FOV, short baselines, see
  Fig. \ref{fig:lonsdale3}).
\item The ionospheric delays differ per station and per source (large FOV,
  long baselines).
\end{enumerate}

The first two regimes can be handled under the self-calibration assumption
used in astronomy that all errors are telescope based thereby reducing the
estimation problem to a single direction independent gain factor per telescope
\cite{Cornwell1994-1}. The problem of estimating these direction independent
gains based on a single calibrator source is treated in \cite{Boonstra2003-1},
while the multiple source case has been discussed in \cite{Wijnholds2006-1}.
The problem of finding a complex gain per antenna per source, the fourth
regime, is not tractable without further assumptions which allow to
parameterize the behavior of the gains over space, time and/or frequency
\cite{Tol2007-1}. In this paper we will focus on the third regime, which is
sketched in Fig. \ref{fig:lonsdale3}, thus filling the gap in the available
literature. It allows for the calibration of individual closely packed groups
of antennas such as a LOFAR station or a subarray of the MWA and PaST
telescopes and thus forms a valuable step towards the calibration of the whole
array.

We will state our problem in general terms, since the problem plays a role in
a range of applications varying from underwater acoustics to antenna arrays.
This explains the persistent interest in on-line calibration, autocalibration
or self-calibration of sensor arrays \cite{Flanagan1999-1, Astely1999-1,
  Pesavento2002-1, See2004-1, Tol2007-1}. In most applications the main driver
for studies on array calibration is to improve the DOA estimation accuracy.
Many studies in this field therefore try to solve for the DOAs and a number of
array parameters. In \cite{Weiss1989-1, See1995-1, Flanagan1999-1}
self-calibration schemes are presented which solve for the direction
independent gains and the sensor positions, i.e. the directional response of
the sensors is assumed known. Other studies assume a controlled environment to
calibrate the array by measuring the calibrator sources one at a time
\cite{Pierre1991-1, Ng1996-1} or exploit the array geometry, e.g. a uniform
linear array (ULA) having a Toeplitz matrix as array covariance matrix
\cite{Astely1999-1, Li1999-1}. Weiss and Friedlander \cite{Weiss1995-1} have
presented a technique for almost fully blind signal estimation. Their work,
however, focuses on estimation and separation of source signals, not on
characterizing the array itself. We thus feel that the problem at hand also
forms an interesting addition to the literature available on sensor array
calibration in general.

In the next section we introduce the data model and provide a mathematical
formulation of the problem. In Sec.\ \ref{sec:analysis} the Cram\`er-Rao bound
for this estimation problem is discussed. Section \ref{sec:algorithms}
presents an alternating least squares (ALS) and a weighted alternating least
squares (WALS) approach optimizing subsets of parameters iteratively.  These
algorithms are validated using Monte Carlo simulations in Sec.\
\ref{sec:simulations}. The conclusions are drawn in Sec.\
\ref{sec:conclusions}.

\emph{Notation}: Overbar $\overline{(\cdot)}$ denotes conjugation, the
transpose operator is denoted by $^T$, the complex conjugate (Hermitian)
transpose by $^H$ and the pseudo-inverse by $^\dagger$.  An estimated value is
denoted by $\expect \{ \cdot \}$. $\odot$ is the element-wise matrix
multiplication (Hadamard product), $\oslash$ is the element-wise matrix
division, $\otimes$ denotes the Kronecker product and $\circ$ is used to
denote the Khatri-Rao or column-wise Kronecker product of two
matrices. $\diag(\cdot)$ converts a vector to a diagonal matrix with the
vector placed on the main diagonal, $\vecdiag(\cdot)$ produces a vector from
the elements of the main diagonal of its argument and $\vec(\cdot)$ converts a
matrix to a vector by stacking the columns of the matrix. We exploit many
properties of Kronecker products, including the following (for matrices and
vectors of compatible dimensions):
\begin{eqnarray}
   \vec(\mathbf{A} \mathbf{B} \mathbf{C}) &=& (\mathbf{C}^T \otimes \mathbf{A})
   \vec(\mathbf{B})
\label{eq:prop1}
\\
   \vec(\mathbf{A} \diag(\mathbf{b}) \mathbf{C}) &=& (\mathbf{C}^T \circ
   \mathbf{A}) \mathbf{b}
\label{eq:prop2}
\\
   (\mathbf{A} \circ \mathbf{B})^H
   (\mathbf{C} \circ \mathbf{D})
   &=&
   \mathbf{A}^H \mathbf{C} \odot \mathbf{B}^H \mathbf{D}
\label{eq:prop3}
\end{eqnarray}

\section {Data model}
\label{sec:datamodel}

We consider an array of $p$ elements located at a single site (``Regime 3'').
Denote the complex baseband output signal of the $i$th array element by
$x_i(t)$ and define the array signal vector $\mathbf{x}(t) = [x_1(t), x_2(t),
\cdots, x_p(t)]^T$. We assume the presence of $q$ mutually independent i.i.d.\
Gaussian signals $s_k(t)$ impinging on the array, which are stacked in a $q
\times 1$ vector $\mathbf{s}(t)$. Likewise the sensor noise signals $n_i(t)$
are assumed to be mutually independent i.i.d.\ Gaussian signals and are
stacked in a $p \times 1$ vector $\mathbf{n}(t)$. If the narrow band condition
holds \cite{Zatman1998-1}, we can define the $q$ spatial signature vectors
$\mathbf{a}_k$ ($p \times 1)$, which describe for the $k$th source
the phase delays at each antenna due only to the geometry.
    
The $q$ sources are considered calibrator sources, thus we assume that their
powers, their nominal positions, and the locations of the antennas (hence
$\mathbf{a}_k$) are known.  Refractive effects caused by ionospheric phase
gradients may shift the apparent locations of the sources, thus we will also
consider cases where the $\mathbf{a}_k$ are only parametrically known.

The sensors are assumed to have the same direction dependent gain behavior
towards the $q$ source signals received by the array.  This can include the
antenna pattern and ionospheric phase effects.  They are described by gain
factors $g_{0k}$ and are collected in a matrix $\mathbf{G_0} = \diag([g_{01},
g_{02}, \cdots, g_{0q}])$. The direction independent gains and phases
(individual receiver gains) can be described as $\bgamma = [\gamma_1,
\gamma_2, \cdots, \gamma_p]^T$ and $\bphi = [\e^{\jcmplx\phi_1},
\e^{\jcmplx\phi_2}, \cdots, \e^{\jcmplx\phi_p}]^T$ respectively with
corresponding diagonal matrix forms $\bGamma = \diag(\bgamma)$ and $\bPhi =
\diag(\bphi)$. With these definitions, the array signal vector can be
described as
\begin{equation}
\mathbf{x}(t) = \bGamma\bPhi \left ( \sum_{k=1}^q \mathbf{a}_k g_{0k} s_k(t)
\right ) + \mathbf{n}(t) = \mathbf{G} \mathbf{A} \mathbf{G}_0 \mathbf{s}(t) +
\mathbf{n}(t)\label{eq:sigvec}
\end{equation}
where $\mathbf{A} = [\mathbf{a}_1, \cdots, \mathbf{a}_k]$ (size $p \times q$)
and $\mathbf{G} = \bGamma \bPhi$; for later use we also define $\bg = \bgamma
\odot \bphi$.

The signal is sampled with period $T$ and $N$ sample vectors are stacked into
a data matrix $\mathbf{X} = [\mathbf{x}(T), \mathbf{x}(2T), \cdots,
\mathbf{x}(NT)]$. The covariance matrix of $\mathbf{x}(t)$ is $\mathbf{R} =
\expect \{ \mathbf{x}(t) \mathbf{x}^H(t) \}$ and is estimated by
$\widehat\mathbf{R} = N^{-1} \mathbf{X} \mathbf{X}^H$. Likewise, the source
signal covariance $\bSigma_s = \diag(\bsigma_s)$ where $\bsigma_s =
[\sigma_{s1}^2, \sigma_{s2}^2, \cdots, \sigma_{sq}^2]^T$ and the noise
covariance matrix is $\bSigma_n = \diag(\bsigma_n)$ where $\bsigma_n =
[\sigma_{n1}^2, \sigma_{n2}^2, \cdots, \sigma_{np}^2]^T$. Then the model for
$\mathbf{R}$ based on \eqref{eq:sigvec} is
\begin{equation}
    \mathbf{R} = \mathbf{G} \mathbf{A} \mathbf{G}_0 \bSigma_s \mathbf{G}_0^H
    \mathbf{A}^H \mathbf{G}^H + \bSigma_n.\label{eq:R}
\end{equation}
In this model, $\bSigma_s$ is known. Since $\mathbf{G}_0$ and $\bSigma_s$ are
diagonal matrices, we can introduce
\begin{eqnarray}
    \bSigma & = & \mathbf{G}_0 \bSigma_s \mathbf{G_0}^H \nonumber\\
    & = & \diag([ |g_{01}|^2\sigma_{s1}^2, \cdots, |g_{0q}|^2\sigma_{sq}^2]) =
    \diag(\bsigma) \,.
\end{eqnarray}
Since the direction dependent gains $g_{0k}$ are not known, the real valued
elements of $\bsigma = [\sigma_1^2, \sigma_2^2, \cdots, \sigma_q^2]^T$ are not
known either. This implies that our problem is identical to estimating an
unknown diagonal signal covariance matrix without any DOA dependent errors. We
may thus restate \eqref{eq:R} as
\begin{equation}
    \mathbf{R} = \mathbf{G} \mathbf{A} \bSigma \mathbf{A}^H \mathbf{G}^H +
    \bSigma_n\label{eq:R2}
\end{equation}
and solve for $\mathbf{G}$, $\bSigma$ and $\bSigma_n$ under the assumption
that $\mathbf{A}$ is known (or parametrically known).

The data model described by \eqref{eq:R2} is commonly used in papers on sensor
array calibration (e.g. \cite{Fuhrmann1994-1, Boonstra2003-1}).  Flanagan and
Bell \cite{Flanagan1999-1}, Weiss and Friedlander \cite{Weiss1989-1} and See
\cite{See1995-1} effectively use the same model, but focus on position
calibration of the array elements and are therefore more explicit on the form
of $\mathbf{A}$. If the source positions and locations of the sensors within
the array are known, an explicit formula for $\mathbf{a}_k$ can be used to
compute the nominal spatial signature vectors.  Estimation of source locations
is required to account for the refractive effects produced by an ionospheric
phase gradient.

The $i$th element of the array is located at $\mathbf{r}_i = [x_i, y_i,
z_i]^T$. These positions can be stacked in a matrix $\mathcal{R} =
[\mathbf{r}_1, \mathbf{r}_2, \cdots, \mathbf{r}_p]^T$ (size $p \times 3$). The
position of the $k$th source can be denoted as $\mathbf{l}_k = [l_k,
m_k, n_k]^T$. The source positions can be stacked in a matrix $\mathcal{L} =
[\mathbf{l}_1, \mathbf{l}_2, \cdots, \mathbf{l}_q]^T$ (size $q \times 3$). The
spatial signature matrix $\mathbf{A}$ can thus be described by
\begin{equation}
\mathbf{A} = \exp \left ( -\jcmplx \frac{2\pi}{\lambda} \mathcal{R}
  \mathcal{L}^T \right )
\end{equation}
where the exponential function is applied element-wise to its argument. In the
remainder of this paper we will specialize to a planar array having $z_i = 0$
for convenience of presentation but without loss of generality.

From \eqref{eq:R2} we observe that $\mathbf{G}$ and $\bSigma$ share a common
scalar factor. Therefore we may impose the boundary condition $\sigma_1^2 =
1$. To solve the ambiguity in the phase solution of $\mathbf{G}$ we will take
the first element as phase reference, i.e. $\phi_1 = 0$ is imposed.\footnote{
  In \cite{Wijnholds2006-2} it is shown that $\sum_{i=1}^p \phi_i = 0$ is the
  optimal constraint for this problem. This constraint has the disadvantage
  that the location of the phase reference is not well defined. Furthermore,
  the choice for the constraint used here simplifies our analysis in
  combination with the constraints required to uniquely identify the source
  locations and the apparent source powers.} When solving for source
locations a similar problem occurs: a single rotation of all DOA vectors can
be compensated by the direction independent gain phase solution. We will
therefore fix the position of the first source.

In this paper we will address four related sensor array calibration problems
based on this data model. These are summarized below where the parameter
vectors adhering to the aforementioned boundary conditions are stated
explicitly.
\begin{enumerate}
\item The sensor noise powers are the same for all elements,
  i.e.\ $\sigma_{n1}^2 = \sigma_{n2}^2 = \cdots = \sigma_{nq}^2$ such that
  $\bSigma_n = \sigma_n^2 \mathbf{I}$ where $\mathbf{I}$ is the identity
  matrix. In this scenario the parameter vector to be estimated is
  $\btheta = [\bgamma^T, \phi_2, \cdots \phi_p, \sigma_2^2, \cdots \sigma_q^2,
  \sigma_n^2]^T$.
\item The sensor noise powers are allowed to differ from one element to the
  other, i.e.\ $\bSigma_n = \diag(\bsigma_n)$. In this case the parameter
  vector is $\btheta = [\bgamma^T, \phi_2, \cdots, \phi_p, \sigma_2^2, \cdots
  \sigma_q^2, \sigma_{n1}^2, \cdots \sigma_{np}^2]^T$.
\item $\bSigma_n = \sigma_n^2 \mathbf{I}$ and $\mathbf{A} =
  \mathbf{A}(\mathcal{L})$, i.e.\ similar to the first scenario but with
  unknown source locations. In this case $\btheta = [\bgamma^T, \phi_2,
  \cdots, \phi_p, \sigma_2^2, \cdots, \sigma_q^2, \sigma_n^2, \mathbf{l}_2^T,
  \cdots, \mathbf{l}_q^T]^T$.
\item $\bSigma_n = \diag(\bsigma_n)$ and $\mathbf{A} =
  \mathbf{A}(\mathcal{L})$, giving $\btheta = [\bgamma^T, \phi_2, \cdots,
  \phi_p, \sigma_2^2, \cdots, \sigma_q^2, \sigma_{n1}^2, \cdots,
  \sigma_{np}^2, \mathbf{l}_2^T, \cdots, \mathbf{l}_q^T]^T$.
\end{enumerate}

\section{CRB Analysis}
\label{sec:analysis}

The Cram\`er-Rao bound (CRB) on the error variance for any unbiased estimator
states that \cite{Kay1993-1}
\begin{equation}
\mathbf{C} = \expect \left \{ \left ( \widehat{\btheta} - \btheta \right )
  \left ( \widehat{\btheta} - \btheta \right )^T \right \} \geq \frac{1}{N}
\mathbf{J}^{-1},
\end{equation}
where $\bC$ is the covariance matrix of $\btheta$, and $\bJ$ is the Fisher
Information Matrix (FIM).  For Gaussian signals the FIM can be expressed as
\begin{equation}
\mathbf{J} = \mathbf{F}^H \left ( \overline{\mathbf{R}}^{-1} \otimes
  \mathbf{R}^{-1} \right ) \mathbf{F}
\end{equation}
where $\mathbf{F}$ is the Jacobian evaluated at the true values of the
parameters, i.e.
\begin{equation}
\mathbf{F} = \frac{\delta \vec ( \mathbf{R} )}{\delta \btheta^T} \Big
  |_{\btheta} \,.
\end{equation}

For the calibration problem defined by the first scenario, the Jacobian can be
partitioned in four parts following the structure in $\btheta$:
\begin{equation}
\mathbf{F} = [\mathbf{F}_\gamma, \mathbf{F}_\phi, \mathbf{F}_\sigma,
\mathbf{F}_{\sigma_n}],
\end{equation}
where it can be shown that
\begin{eqnarray}
\mathbf{F}_\gamma & = & \left ( \overline{\mathbf{G} \mathbf{R}_0} \bPhi
\right ) \circ \mathbf{I} + \mathbf{I} \circ \left ( \mathbf{G} \mathbf{R}_0
  \overline{\bPhi} \right )\\
\mathbf{F}_\phi & = & \jcmplx \left ( \left ( \overline{\mathbf{G}
      \mathbf{R}_0} \mathbf{G} \right ) \circ \mathbf{I} - \mathbf{I} \circ
  \left ( \mathbf{G} \mathbf{R}_0 \overline{\mathbf{G}} \right ) \right )
\mathbf{I}_s\\
\mathbf{F}_\sigma & = & \left ( \left ( \overline{\mathbf{G} \mathbf{A}}
  \right ) \circ \left ( \mathbf{G} \mathbf{A} \right ) \right )
\mathbf{I}_s\\
\mathbf{F}_{\sigma_n} & = & \vec ( \mathbf{I} ) \label{eq:Fsigman}
\end{eqnarray}
$\mathbf{I}_s$ is a selection matrix of appropriate size equal to the identity
matrix with its first column removed so that the derivatives with respect to
$\phi_1$ and $\sigma_1^2$ are omitted.

In the second scenario the Jacobian can be partitioned in a similar way, the
difference being that the expression for $\mathbf{F}_{\sigma_n}$ given by
Eq. \eqref{eq:Fsigman} should be replaced by
\begin{equation}
\mathbf{F}_{\sigma_n} = \mathbf{I} \circ \mathbf{I}
\end{equation}

In the third and fourth scenarios an additional component
$\mathbf{F}_\mathbf{l}$ is added to the Jacobian containing the derivatives of
$\vec(\mathbf{R})$ with respect to the source position coordinates. This
component can be partitioned as $\mathbf{F}_\mathbf{l} = [\mathbf{F}_l,
\mathbf{F}_m]$. Introducing
\begin{eqnarray}
\mathbf{G}_x & = & \diag([x_1, x_2, \cdots x_p]^T) \mathbf{G}\\
\mathbf{G}_y & = & \diag([y_1, y_2, \cdots y_p]^T) \mathbf{G}
\end{eqnarray}
these components can be conveniently written as
\begin{eqnarray}
\mathbf{F}_l & = & -\jcmplx \frac{2\pi}{\lambda} \left ( \overline{\mathbf{G}
    \mathbf{A}} \circ \mathbf{G}_x \mathbf{A} - \overline{\mathbf{G}_x
    \mathbf{A}} \circ \mathbf{G} \mathbf{A} \right ) \bSigma \mathbf{I}_s\\
\mathbf{F}_m & = & -\jcmplx \frac{2\pi}{\lambda} \left ( \overline{\mathbf{G}
    \mathbf{A}} \circ \mathbf{G}_y \mathbf{A} - \overline{\mathbf{G}_y
    \mathbf{A}} \circ \mathbf{G} \mathbf{A} \right ) \bSigma \mathbf{I}_s
\end{eqnarray}

These equations show that the entries of the Jacobian related to derivatives
with respect to the $l$- and $m$-coordinates of the sources are proportional
to the $x$- and $y$-coordinates of the array elements respectively. The
physical interpretation of this relation is that a plane wave propagating
along the coordinate axis of the coordinate to be estimated provides a more
useful test signal to estimate the source location than a signal propagating
perpendicular to this axis.

\section{Algorithms}
\label{sec:algorithms}

\subsection{Generalized least squares formulation}

An asymptotically efficient estimate of the model parameters $\btheta$ can be
obtained via the ML formulation. Since all signals are assumed to be i.i.d.
Gaussian signals, the derivation is standard and ML parameter estimates for
$N$ independent samples are obtained by minimizing the negative
$\log$-likelihood function \cite{Ottersten1998-1}
\begin{equation}
\widehat\btheta = \underset{\btheta}{\mathrm{argmin}} \left ( \ln \left |
  \mathbf{R}(\btheta) \right | + \mathrm{tr} \left ( \mathbf{R}^{-1}(\btheta)
  \widehat\mathbf{R} \right ) \right )
\end{equation}
where $\mathbf{R}(\btheta)$ is the model covariance matrix as function of
$\btheta$ and $\widehat\mathbf{R}$ is the sample covariance matrix $N^{-1}
\mathbf{X}\mathbf{X}^H$.

It does not seem possible to solve this minimization problem in closed
form. As discussed in \cite{Ottersten1998-1} a weighted least squares
covariance matching approach is known to lead to estimates that are, for a
large number of samples, equivalent to ML estimates and are therefore
asymptotically efficient and reach the CRB.

The least squares covariance model fitting problem can be defined as
\begin{equation}
\widehat{\btheta} = \underset{\btheta}{\mathrm{argmin}} \left \Arrowvert
  \widehat\mathbf{R} - \mathbf{R}(\btheta) \right \Arrowvert_F^2.
\end{equation}
Equivalently, we consider minimization of the cost function
\begin{equation}
\kappa(\btheta) = \left \Arrowvert \widehat\mathbf{R} - \mathbf{R}(\btheta)
 \right \Arrowvert_F^2 = \mathbf{f}^H(\btheta) \mathbf{f}(\btheta)
\end{equation}
where $\mathbf{f}(\btheta) = \vec \left ( \widehat\mathbf{R} -
  \mathbf{R}(\btheta) \right )$. The more general weighted least squares
problem is obtained by introducing a weighting matrix $\mathbf{W}$ and
optimizing
\begin{equation}
\kappa_\mathbf{W}(\btheta) = \mathbf{f}^H(\btheta) \mathbf{W}
\mathbf{f}(\btheta).
\end{equation}
The optimal weight is known to be the inverse of the asymptotic covariance of
the residuals, $\expect \left \{ \mathbf{f}(\btheta_0) \mathbf{f}(\btheta_0)^H
\right \}$, where $\btheta_0$ is the true value of the parameters
\cite{Ottersten1998-1}. The optimal weight for Gaussian sources is thus
\begin{equation}
\mathbf{W}_{opt} = \left (\overline{\mathbf{R}} \otimes \mathbf{R} \right
)^{-1} = \overline{\mathbf{R}}^{-1} \otimes \mathbf{R}^{-1}.
\end{equation}
The Kronecker structure of $\mathbf{W}_{opt}$ allows to introduce
$\mathbf{W}_c = \mathbf{R}^{-\frac{1}{2}}$ and write the weighted least
squares (WLS) cost function as
\begin{equation}
\kappa_\bW (\btheta) = \left \Arrowvert\mathbf{W}_c \left (
    \widehat\mathbf{R} - \mathbf{R} (\btheta) \right ) \mathbf{W}_c \right
\Arrowvert_F^2.
\label{eq:weightcost}
\end{equation}
As mentioned, this estimator is asymptotically unbiased and asymptotically
efficient \cite{Ottersten1998-1}.

We propose to solve the least squares problems for the four cases defined in
section \ref{sec:datamodel} by alternating between least squares solutions to
subsets of parameters. The subsets are chosen such that the solution is either
available in the literature or can be derived analytically.  We will have four
subsets of parameters: the complex direction independent gains of the elements
$\mathbf{g}$, the apparent source powers $\bsigma$, the receiver noise powers
$\bsigma_n$, and the source locations $\mathcal{L}$. In the following
subsections we will develop the unweighted and weighted least squares
solutions for these subsets before putting them together to form the
alternating least squares (ALS) or weighted alternating least squares (WALS)
method respectively.

\subsection{Estimation of direction independent gains}

The least squares problem to find the omnidirectional gains $\widehat{\bg}$
based on the weighted cost function $\kappa_\bW \left ( \btheta \right )$ can
be formulated as
\begin{eqnarray}
\widehat{\bg} & = & \underset{\bg}{\argmin} \Big \Arrowvert \left (
    \overline{\bW}_c \otimes \bW_c \right ) \vec \left ( \widehat{\bR} -
    \bSigma_n \right ) + \nonumber\\
& & - \left ( \overline{\bW}_c \otimes \bW_c \right ) \diag \left ( \vec \left
    ( \bR_0 \right ) \right ) \left ( \overline{\bg} \otimes \bg \right ) \Big
\Arrowvert_F^2
\end{eqnarray}
where we have introduced $\bR_0 = \bA \bSigma \bA^H$; in this subsection,
$\bR_0$ is assumed to be known. This problem can be solved using standard
techniques by regarding $\bg$ and $\overline{\bg}$ as independent vector
parameters and alternatingly solve for them until convergence\footnote{We
  thank one of the reviewers for pointing this out.}. In this approach, the
solution for $\bg$ is
\begin{eqnarray}
\widehat{\bg} & = & \underset{\bg}{\mathrm{argmin}} \Big \Arrowvert \left (
    \overline{\bW}_c \otimes \bW_c \right ) \vec \left ( \widehat{\bR}
      - \bSigma_n \right ) + \nonumber\\
& & - \left ( \overline{\bW}_c \otimes \bW_c \right ) \left ( \left ( \bR_0
    \bG^H \right ) \circ \bI \right ) \bg \Big \Arrowvert_F^2 \nonumber\\
& = & \left ( \overline{\bR_0^H \bG^H \bR^{-1} \bG \bR_0} \odot \bR^{-1}
  \right )^{-1} \times \nonumber\\
& & \left ( \overline{\bR^{-1} \bG \bR_0} \circ \bR^{-1} \right )^H \vec \left
  ( \widehat{\bR} - \bSigma_n \right ). \label{eq:glin}
\end{eqnarray}
Since the result for $\overline{\bg}$ is simply the complex conjugate of this
relation, it is sufficient to apply Eq.\ \eqref{eq:glin} repeatedly until
convergence. Although ALS ensures that the value of the cost function
decreases in each iteration, it does not guarantee convergence to the global
minimum, especially if the initial estimate is poor. The number of iterations
required for convergence also depends strongly on the vicinity of the initial
estimate to the true value. We are thus interested in obtaining a good initial
estimate for $\bg$.

Fuhrmann \cite{Fuhrmann1994-1} has proposed to use a suboptimal closed form
solution to initialize the Newton iterations used to solve the ML cost
function under the assumption that $\bSigma_n$ is known (or actually no noise
is present). A similar problem (with unknown $\bSigma_n$ and $\mathbf{R}_0$ a
rank 1 matrix, i.e.\ a single calibrator source) was also studied by us in
\cite{Boonstra2003-1}, where a ``column ratio method'' (COLR) was
proposed. Below, we generalize and improve that technique\footnote{An initial
  presentation of this was in \cite{Wijnholds2006-1}.}. All these techniques
are suboptimal in the case of multiple calibrator sources, but they can be
used to provide the starting point for iterative refinement.

The closed form solution by \cite{Fuhrmann1994-1} is derived as follows. If
the structure in $\mathbf{v} = \overline{\mathbf{g}} \otimes \mathbf{g}$ is
neglected and $\bSigma_n$ is known or negligible, we can solve for
$\mathbf{v}$ in the least squares sense by
\begin{equation}
\widehat{\mathbf{v}} = \diag \left ( \vec \left ( \mathbf{R}_0 \right ) \right
)^{-1} \vec \left ( \widehat{\mathbf{R}} - \bSigma_n \right )
\label{eq:vecggh}
\end{equation}
or equivalently
\begin{equation}
\widehat{\mathbf{g}\mathbf{g}^H} = \left ( \widehat{\mathbf{R}} - \bSigma_n
\right ) \oslash \mathbf{R}_0.
\end{equation}
Note that the weights cancel since we solve for one parameter for each entry
of $\widehat{\bR}$. Subsequently, use the structure of $\mathbf{v}$ and assume
that the estimate of $\mathbf{g}\mathbf{g}^H$ obtained in the first step has
rank 1. An eigenvalue decomposition is used to extract $\widehat{\mathbf{g}}$
as the dominant eigenvector from $\widehat{\mathbf{v}}$.  It is clear however
that the noise on specific elements of $\widehat{\mathbf{v}}$ may be increased
considerably if some entries of $\mathbf{R}_0$ have a small value. Also, the
method is not applicable if $\bSigma_n$ is unknown.

In \cite{Wijnholds2006-1} another approach is therefore suggested based on the
observation that $g_i \overline{g}_k = R_{ik} / R_{0,ik}$ holds for all
off-diagonal elements of $\mathbf{R}$, i.e., for $i \neq k$.  This implies
that
\begin{equation}
\frac{g_i}{g_j} = \frac{g_i \overline{g}_k}{g_j \overline{g}_k} = \frac{R_{ik}
  / R_{0,ik}}{R_{jk} / R_{0,jk}} = \frac{R_{ik} R_{0,jk}}{R_{0,ik}
  R_{jk}}
  \quad (k \neq i,j).
  \label{eq:gratio}
\end{equation}
This relation is similar to the closure amplitude relation in astronomy
\cite{Cornwell1994-1,Pearson1984-1}, which states that in an observation on a
{\em single} point source or, more generally, in the case of a rank 1 model,
the amplitude ratios are related as indicated by the second equality sign in
\eqref{eq:gratio}.

Since the index $k$ can be chosen freely as long as $k \neq i,j$, we can
introduce $\mathbf{c}_{1,ij}$ being the column vector containing the values
$R_{ik} R_{0,jk}$ and $\mathbf{c}_{2,ij}$ being the column vector containing
the values $R_{0,ik} R_{jk}$ for all possible values of $k \neq i,j$.  We can
now write \eqref{eq:gratio} in its more general form
\begin{equation}
\frac{g_i}{g_j} \mathbf{c}_{2,ij} = \mathbf{c}_{1,ij},
\label{eq:gratiogeneral}
\end{equation}
which has the well-known solution
\begin{equation}
\frac{g_i}{g_j} = \mathbf{c}_{2,ij}^\dagger \mathbf{c}_{1,ij}
\label{eq:gratiopinv}.
\end{equation}

All possible gain ratios can be collected in a matrix $\mathbf{M}$ with
entries $M_{ij} = \mathbf{c}_{2,ij}^\dagger \mathbf{c}_{1,ij}$. Since the
model for $\mathbf{M}$ is $M_{ij} = g_i/g_j$, the matrix $\mathbf{M}$ is
expected to be of rank 1, and $\mathbf{g}$ can be extracted from this matrix
using an eigenvalue decomposition: let $\bu_1$ be the eigenvector
corresponding to the largest eigenvalue of $\bM$, then $\bg = \alpha \bu_1$,
where the scaling $\alpha$ needs to be determined separately since the
quotient $g_i/g_j$ is insensitive to modification by a constant scaling factor
applied to all gains. This scaling can be found by minimizing
\begin{equation}
{\widehat{\alpha}} = \underset{\alpha}{\mathrm{argmin}} \parallel \alpha
\mathbf{G} \mathbf{R}_0 \mathbf{G}^H \overline{\alpha} - \widehat{\mathbf{R}}
\parallel_F^2
\end{equation}
for all off-diagonal elements. By introducing the vectors $\mathbf{r}_0 =
\vec_-(\mathbf{G} \mathbf{R}_0 \mathbf{G}^H)$ and $\mathbf{r} =
\vec_-(\widehat{\mathbf{R}})$, where $\vec_-(\cdot)$ operates like the
$\vec(\cdot)$ operator but leaves out the elements on the main diagonal of
this argument, this cost function can rewritten as
\begin{equation}
{\widehat{\alpha}} = \underset{\alpha}{\mathrm{argmin}} \parallel |\alpha|^2
\mathbf{r}_0 - \mathbf{r} \parallel_F^2 \label{eq:alpha}
\end{equation}
which can be solved in the least squares sense by $\widehat{\alpha} =
\sqrt{\mathbf{r}_0^\dagger \mathbf{r}}$.

Equation \eqref{eq:gratiopinv} extends the column ratio method
proposed in \cite{Boonstra2003-1} to all elements of the matrix. The column
ratio method was introduced to reconstruct the main diagonal of a rank 1
source model, which allows to estimate this rank 1 model using an
eigenvalue decomposition without
distortion of the results due to unknown sensor noise powers. In our case this
allows us to neglect the unknown receiver noise powers. The CRB analysis in
\cite{Tol2006-1} shows that it does not matter whether one simultaneously
estimates the receiver noise powers and the omnidirectional complex gains
exploiting all data, i.e. including the autocorrelations, or ignore the
autocorrelations and solve for the direction independent complex gains
only.

The Monte Carlo simulations presented in \cite{Wijnholds2006-1} suggest that
the method outlined above provides a statistically efficient estimate of
$\widehat{\mathbf{g}}$; the pseudo-inverse ensures that the noise on the
entries of $\mathbf{M}$ does not increase dramatically due to small values in
either $\widehat{\mathbf{R}}$ or $\mathbf{R}_0$. It also provides a
modification of the initial estimate proposed by Fuhrmann
\cite{Fuhrmann1994-1} which finds a near optimal solution to the least squares
gain estimation problem without further optimization using, e.g., the Newton
algorithm. This will generally save computational effort. In the simulations
in this paper, we will use Eq.\ \eqref{eq:gratiopinv} to demonstrate that this
method gives statistically efficient results without requiring additional
iterations using Eq.\ \eqref{eq:glin}. Moreover, if three or more iterations
are required to ensure convergence using Eq.\ \eqref{eq:glin}, it also
requires less computational effort, as discussed in Sec.\
\ref{ssec:complexity}.

Although the pseudo-inverse in Eq. \eqref{eq:gratiopinv} ensures robustness
against small entries of either $\widehat{\mathbf{R}}$ or $\mathbf{R}_0$, the
fact that the method relies on gain ratios may lead to poor performance if
there are small entries in $\bg$, e.g. due to failing array elements. This
risk can be mitigated by rewriting Eq. \eqref{eq:gratiogeneral} to
\begin{equation}
\bc_{2,ij} g_i = \bc_{1,ij}  g_j.
\end{equation}
By defining
\begin{eqnarray}
\by_i & = & \left [ \bc_{2,i1}^T, \bc_{2,i2}^T, \cdots, \bc_{2,ip}^T \right
]^T \nonumber\\
\bY_i & = & \left [ \begin{array}{cccc} \bc_{1,i1} & & &\\ & \bc_{1,i2} & &\\
    & & \ddots &\\ & & & \bc_{1,ip} \end{array} \right ] \nonumber\\
\bC_1 & = & \left [ \bY_1^T, \bY_2^T, \cdots, \bY_p^T \right ]^T \nonumber\\
\bC_2 & = & \left [ \begin{array}{cccc} \by_1 & & &\\ & \by_2 & &\\ & &
    \ddots &\\ & & & \by_p \end{array} \right ] \nonumber
\end{eqnarray}
we can enumerate $i$ and $j$ and collect all relations in a single matrix
equation
\begin{equation}
\bC_2 \bg = \bC_1 \bg.
\label{eq:gainequation}
\end{equation}

This suggests that $\bg$ can be found by searching for the null space of the
$p(p-1)^2 \times p$ matrix $\bC_2 - \bC_1$. By substituting the details of
$\bc_{1,ij}$ and $\bc_{2,ij}$ in $\bC_2 - \bC_1$, it is easily seen that $\bg$
lies indeed in the null space and that this null space is one
dimensional. However, this is only true for noise free data. In the practical
case of noisy data, there will not be a null space and $\bg$ lies in the noise
subspace. Furthermore, finding $\bg$ will involve a singular value
decomposition on a very large matrix which may be computationally prohibitive.

An alternative approach follows from recognizing that $\bg$ is obtained from
the principal right eigenvector of
\begin{eqnarray}
\bC_2^\dagger \bC_1 & = & \left [ \frac{\sum_k
    \overline{R}_{0,ik} \overline{R}_{jk} R_{0,jk} R_{ik}}{\sum_j \sum_k
    \overline{R}_{0,ik} \overline{R}_{jk} R_{0,ik} R_{jk}} \right ]
\nonumber\\
& = & \left [ \frac{g_i \overline{g}_j \sum_k \left | g_k \right |^2 \left |
      R_{0,ik} \right |^2 \left | R_{0,jk} \right |^2}{\sum_j \left | g_j
    \right |^2 \sum_k \left | g_k \right |^2 \left | R_{0,ik} \right |^2 \left
      | R_{0,jk} \right |^2} \right ]\label{eq:pinvC2C1},
\end{eqnarray}
which is easily found by substitution of the definition of $\bc_{1,ij}$ and
$\bc_{2,ij}$ and use of the Moore-Penrose left inverse. Equation
\eqref{eq:pinvC2C1} shows that $\bg$ is the principal right eigenvector of
$\bC_2^\dagger \bC_1$, which has eigenvalue 1 in the noise free
case. Simulations with completely randomized $\bR_0$ and $\bg$ suggest that
the other eigenvalues are considerably lower than 1. Since it is easily
demonstrated that the noise on actual data will lower the main eigenvalue,
this is an important result; the contrast between the largest and the
second-largest eigenvalue determines the susceptibility of this method to
noise on the data.

Note that these methods are still insensitive to a constant scaling factor
applied to all gains. This ambiguity is solved by Eq.\ \eqref{eq:alpha}. Also
note that all methods presented above enumerate over $i, j \neq k$ to avoid
the diagonal entries of the array covariance matrix which are affected by the
system noise. By imposing more stringent restrictions on the choice of $i$ and
$j$, this method can also handle cases in which some elements in the array
experience correlator noise, i.e. cases in which $\bSigma_n$ is not diagonal
but still has sufficient zero entries to allow the above methods to extract
the required information for every combination of $g_i$ and $g_j$.

\subsection{Source power estimation}

Based on the unweighted cost function $\kappa(\btheta)$, $\widehat\bsigma$ is
found by
\begin{eqnarray}
  \widehat\bsigma & = & \underset{\bsigma}{\mathrm{argmin}} \left \Arrowvert
    \widehat\mathbf{R} - \mathbf{G} \mathbf{A} \bSigma \mathbf{A}^H
    \mathbf{G}^H - \bSigma_n \right \Arrowvert_F^2 \nonumber\\
  & = & \underset{\bsigma}{\mathrm{argmin}} \left \Arrowvert \vec \left (
      \left ( \widehat\mathbf{R} - \bSigma_n \right ) - \left ( \mathbf{G}
        \mathbf{A} \right ) \bSigma \left ( \mathbf{G} \mathbf{A} \right )^H
    \right ) \right \Arrowvert_F^2 \nonumber\\
  & = & \underset{\bsigma}{\mathrm{argmin}} \left \Arrowvert \vec \left (
      \widehat\mathbf{R} - \bSigma_n \right ) - \left ( \overline{\bG}
      \overline{\bA} \circ \bG \bA \right ) \bsigma \right \Arrowvert_F^2
  \nonumber\\
  & = & \left ( \overline{\mathbf{G} \mathbf{A}} \circ \mathbf{G} \mathbf{A}
  \right )^\dagger \vec \left ( \widehat\mathbf{R} - \bSigma_n \right )
\label{eq:sigma}
\end{eqnarray}

If weighting is applied, $\widehat\bsigma$ follows from
\begin{eqnarray}
\widehat\bsigma & = & \underset{\bsigma}{\mathrm{argmin}} \Big \Arrowvert
\mathbf{W}_c \left ( \widehat\mathbf{R} - \bSigma_n \right ) \mathbf{W}_c -
\nonumber\\
& & \mathbf{W}_c \mathbf{G} \mathbf{A} \bSigma \mathbf{A}^H \mathbf{G}^H
\mathbf{W}_c \Big \Arrowvert_F^2 \nonumber\\
& = & \underset{\bsigma}{\mathrm{argmin}} \Big \Arrowvert \left (
  \overline{\mathbf{W}}_c \otimes \mathbf{W}_c \right ) \vec \left (
  \widehat\mathbf{R} - \bSigma_n \right ) - \nonumber\\
& & \left ( \overline{\mathbf{W}_c \mathbf{G} \mathbf{A}} \right ) \circ \left
  ( \mathbf{W}_c \mathbf{G} \mathbf{A} \right ) \bsigma_n \Big \Arrowvert_F^2
\nonumber\\
& = & \left ( \left ( \overline{\mathbf{W}_c \mathbf{G} \mathbf{A}} \right )
  \circ \left ( \mathbf{W}_c \mathbf{G} \mathbf{A} \right ) \right )^\dagger
\times \nonumber\\
& & \left ( \overline{\mathbf{W}}_c \otimes \mathbf{W}_c \right ) \vec \left (
  \widehat{\mathbf{R}} - \bSigma_n \right ).
\end{eqnarray}
Using standard relations for Khatri-Rao and Kronecker products and
substituting $\mathbf{W}_c = \mathbf{R}^{-\frac{1}{2}}$ we obtain
\begin{eqnarray}
\widehat\bsigma & = & \left ( \left ( \overline{\mathbf{A}^H \mathbf{G}^H
      \mathbf{R}^{-1} \mathbf{G} \mathbf{A}} \right ) \odot \left (
    \mathbf{A}^H \mathbf{G}^H \mathbf{R}^{-1} \mathbf{G} \mathbf{A} \right )
\right )^{-1} \times \nonumber\\
& & \vecdiag \left ( \mathbf{A}^H \mathbf{G}^H \mathbf{R}^{-1} \left (
    \widehat\mathbf{R} - \bSigma \right ) \mathbf{R}^{-1} \mathbf{G}
  \mathbf{A} \right ) \label{eq:sigmaw}
\end{eqnarray}
This result confirms the observation by Ottersten, Stoica and Roy
\cite{Ottersten1998-1} that although the derivation involves a square root of
the array covariance matrix, the final result only depends on the array
covariance matrix and its inverse.

\subsection{Estimating receiver noise powers}

If $\bSigma_n = \diag \left ( \bsigma_n \right )$ and no weighting is applied
to the cost function, then $\widehat\bsigma_n$ is found by solving
\begin{equation}
\widehat\bsigma_n = \underset{\bsigma_n}{\mathrm{argmin}} \left \Arrowvert
\widehat\mathbf{R} - \mathbf{G} \mathbf{A} \bSigma \mathbf{A}^H \mathbf{G}^H -
\bSigma_n \right \Arrowvert_F^2.
\end{equation}
This estimation problem is the same as the sensor noise estimation problem
treated in \cite{Boonstra2003-1}, so we just state the result, which is found
to be
\begin{equation}
\widehat\bsigma_n = \diag \left ( \widehat\mathbf{R} - \mathbf{G} \mathbf{A}
  \bSigma \mathbf{A}^H \mathbf{G}^H \right ).\label{eq:sigman1}
\end{equation}
If $\bSigma_n = \sigma_n^2 \mathbf{I}$, a similar derivation gives
\begin{equation}
\widehat\sigma_n^2 = \frac{1}{p} \trace \left ( \widehat\mathbf{R} -
  \mathbf{G} \mathbf{A} \bSigma \mathbf{A}^H \mathbf{G}^H \right
).\label{eq:sigman2}
\end{equation}
This result is just the average of the sensor noise estimates obtained when
they are estimated individually.

If the weighted cost function $\kappa_\mathbf{W}(\btheta)$ is used, we can
follow a derivation similar to the one that led us to the estimate for
$\widehat{\bsigma}$ in the previous section. We first note that
\begin{eqnarray}
\widehat\bsigma_n & = & \underset{\bsigma_n}{\mathrm{argmin}} \Big \Arrowvert
  \mathbf{W}_c \left ( \widehat\mathbf{R} - \mathbf{G} \mathbf{R}_0
    \mathbf{G}^H \right ) \mathbf{W}_c - \nonumber\\
& & \mathbf{W}_c \bSigma_n \mathbf{W}_c \Big \Arrowvert_F^2
\nonumber\\
& = & \underset{\bsigma_n}{\mathrm{argmin}} \Big \Arrowvert \left (
  \overline{\mathbf{W}}_c \otimes \mathbf{W}_c \right ) \vec \left (
  \widehat\mathbf{R} - \mathbf{G} \mathbf{R}_0 \mathbf{G}^H
\right ) - \nonumber\\
& & \left ( \overline{\mathbf{W}}_c \circ \mathbf{W}_c \right ) \bsigma_n
\Big \Arrowvert_F^2 \nonumber\\
& = & \left ( \overline{\mathbf{W}}_c \circ \mathbf{W}_c \right )^\dagger
\left ( \overline{\mathbf{W}}_c \otimes \mathbf{W}_c \right ) \vec \left (
  \widehat\mathbf{R} - \mathbf{G} \mathbf{R}_0 \mathbf{G}^H \right
). \nonumber\\
& &
\end{eqnarray}
Applying the standard Khatri-Rao and Kronecker product relations and inserting
$\mathbf{W}_c = \mathbf{R}^{-\frac{1}{2}}$ we get
\begin{eqnarray}
\widehat\bsigma_n & = & \left ( \overline{\mathbf{R}}^{-1} \odot
  \mathbf{R}^{-1} \right )^{-1} \times \nonumber\\
& & \vecdiag \left ( \mathbf{R}^{-1} \left ( \widehat\mathbf{R} - \mathbf{G}
    \mathbf{R}_0 \mathbf{G}^H \right ) \mathbf{R}^{-1} \right
).\label{eq:sigman1w}
\end{eqnarray}

A similar derivation for $\bSigma_n = \sigma_n^2 \mathbf{I}$ gives
\begin{equation}
\widehat\sigma_n^2 = \Arrowvert \mathbf{R}^{-1} \Arrowvert_F^{-2} \trace \left
  ( \mathbf{R}^{-1} \left ( \widehat\mathbf{R} - \mathbf{G} \mathbf{R}_0
    \mathbf{G}^H \right ) \mathbf{R}^{-1} \right ). \label{eq:sigman2w}
\end{equation}

The true value of the covariance matrix $\mathbf{R}$ is not known in practical
situations. Therefore the measured covariance matrix $\widehat{\mathbf{R}}$ is
generally taken as estimate of $\mathbf{R}$. It can be shown that this
conventional approach leads to a noticeable bias in the estimate of
$\widehat\bsigma_n$ for a finite number of samples $N$. For simplicity of the
argument we will prove this statement for the case where $\bR = \sigma_n^2
\mathbf{I}$, such that $\mathbf{G} \mathbf{R}_0 \mathbf{G}^H = 0$. Inserting
this in Eq. \eqref{eq:sigman2w} gives
\begin{eqnarray}
\widehat{\sigma}_n^2 & = & \parallel \widehat{\mathbf{R}}^{-1}
\parallel_F^{-2} \trace \left ( \widehat{\mathbf{R}}^{-1} \widehat{\mathbf{R}}
  \widehat{\mathbf{R}}^{-1} \right ) \nonumber\\
& = & \parallel \widehat{\mathbf{R}}^{-1} \parallel_F^{-2} \trace \left (
  \widehat{\mathbf{R}}^{-1} \right ). \label{eq:biassigman1}
\end{eqnarray}
Since $\mathbf{R}^{-1} = \sigma_n^{-2} \mathbf{I}$, we can write
$\widehat{\mathbf{R}}^{-1} = \sigma_n^{-2} \left ( \mathbf{I} + \mathbf{E}
\right )$ where $\sigma_n^{-2} \mathbf{E} = \widehat{\mathbf{R}}^{-1} -
\mathbf{R}^{-1}$ represents a specific realization of the noise on the
elements of $\widehat{\mathbf{R}}^{-1}$. After substitution in
Eq. \eqref{eq:biassigman1} we obtain
\begin{eqnarray}
\widehat{\sigma}_n^2 & = & \parallel \sigma_n^{-2} \left ( \mathbf{I} +
  \mathbf{E} \right ) \parallel_F^{-2} \trace \left ( \sigma_n^{-2} \left (
    \mathbf{I} + \mathbf{E} \right ) \right ) \nonumber\\
& = & \sigma_n^2 \trace \left ( \left ( \mathbf{I} + \mathbf{E} \right ) \left
    ( \mathbf{I} + \mathbf{E} \right )^H \right )^{-1} \trace \left (
  \mathbf{I} + \mathbf{E} \right ) \nonumber\\
& = & \sigma_n^2 \left ( \trace \left ( \mathbf{I} \right ) + \trace \left (
    \mathbf{E} \right ) + \trace \left ( \mathbf{E}^H \right ) + \trace \left
    ( \mathbf{E} \mathbf{E}^H \right ) \right )^{-1} \times \nonumber\\
& & \times \left ( \trace \left ( \mathbf{I} \right ) + \trace \left (
    \mathbf{E} \right ) \right ).
\end{eqnarray}
Since $\mathbf{E}$ represents the noise on the data, the expected value of
$\trace \left ( \mathbf{E} \right )$ and $\trace \left ( \mathbf{E}^H \right
)$ is zero. This implies that
\begin{eqnarray}
\widehat{\sigma}_n^2 & \approx & \sigma_n^2 \left ( \trace \left ( \mathbf{I}
  \right ) + \trace \left ( \mathbf{E} \mathbf{E}^H \right ) \right )^{-1}
\trace \left ( \mathbf{I} \right ) \nonumber\\
& = & \sigma_n^2 \frac{\parallel \mathbf{I} \parallel_F^2}{\parallel
  \mathbf{I} \parallel_F^2 + \parallel \mathbf{E} \parallel_F^2}.
\end{eqnarray}
    This shows that the presence of noise systematically lowers the value
    of the estimate of $\sigma_n^2$. Since $\widehat{\mathbf{R}}^{-1}$
    asymptotically in $N$ converges to $\mathbf{R}^{-1}$, $\mathbf{E}$
    converges to zero. Therefore this result also shows that the
    $\sigma_n^2$ estimate asymptotically converges to the true value if the
    number of samples $N$ approaches infinity.

    However, this bias can be avoided by using the best available
    knowledge of the estimated parameters, i.e., in (\ref{eq:sigman2w}) use
\begin{equation}
\mathbf{R} = \widehat\bGamma \mathbf{A} \left ( \widehat{\mathcal{L}} \right )
\bSigma \mathbf{A}^H \left ( \widehat{\mathcal{L}} \right ) \widehat \bGamma^H
+ \widehat{\bSigma}_n.
\end{equation}
In the first iteration of an alternating least squares algorithm, initial
estimates of the parameters are used. These values are replaced by
increasingly accurate estimated values in consecutive iterations.

\subsection{DOA estimation}
\label{subsec:WSF}

    The problem of estimating $\cal{L}$ is that of estimating the direction of
    arrival of signals impinging on a sensor array, and has been studied
    extensively. MUSIC \cite{Schmidt1986-1} and weighted subspace fitting
    (WSF) \cite{Viberg1991-1, Viberg1991-2} are well-known statistically
    efficient DOA estimation methods applicable to arbitrary sensor
    arrays. In either case the eigenvalue decomposition of $\mathbf{R}$
    is interpreted in terms of a noise subspace and a signal subspace, i.e.
\begin{equation}
\mathbf{R} = \sum_{i=1}^p \lambda_i \mathbf{e}_i\mathbf{e}_i^H = \mathbf{E}_s
\bLambda_s \mathbf{E}_s^H + \mathbf{E}_n \bLambda_n \mathbf{E}_n^H
\end{equation}
    where $\lambda_1 > \cdots > \lambda_q > \lambda_{q+1} \geq \cdots \geq
    \lambda_p$.
    The noise eigenvalues of the whitened true covariance matrix are all
    equal and the number of sources can be derived from the distribution of
    the eigenvalues of $\mathbf{R}$; we will assume that it is known. In
    practical situations in which only an estimate of the true covariance
    matrix is available, methods like the exponential fitting test
    \cite{Quinlan2006-1, Quinlan2007-1} or information theoretic criteria
    \cite{Wax1985-1} must be used to determine the number of signals.

    It is quite straightforward to adapt the WSF method to our needs. The data
    model assumed in \cite{Viberg1991-2} can be described as
\begin{equation}
\mathbf{R} = \widetilde{\mathbf{A}} \left ( \mathcal{L} \right ) \bSigma
\widetilde{\mathbf{A}}^H \left ( \mathcal{L} \right ) + \sigma_n^2 \mathbf{I}
\label{eq:modelWSF}
\end{equation}
where we have used $\widetilde{\mathbf{A}}$ to avoid confusion with
$\mathbf{A}$ introduced in this paper. To map our data model on the data model
described by Eq. \eqref{eq:modelWSF} we should whiten our array covariance
matrix using
\begin{equation}
\mathbf{R}_w = \bSigma_n^{-\frac{1}{2}} \mathbf{R} \bSigma_n^{-\frac{1}{2}} =
\bSigma_n^{-\frac{1}{2}} \mathbf{G} \mathbf{A} \bSigma \mathbf{A}^H
\mathbf{G}^H \bSigma_n^{-\frac{1}{2}} + \mathbf{I}
\end{equation}
and then set $\widetilde{\mathbf{A}} = \bSigma_n^{-1/2} \mathbf{G}
\mathbf{A}$. With these substitutions it is straightforward to implement the
procedures described in \cite{Viberg1991-2}.

\subsection{Alternating Least Squares (ALS) and Weighted Alternating Least Squares (WALS)}

The ingredients of the previous four subsections can be combined to formulate
a (Weighted) Alternating Least Squares solution to the stated optimization
problems.  We start by introducing an algorithm handling the first two
scenarios identified in section \ref{sec:datamodel}. DOA estimation is then
added in a straightforward way.

To estimate $\widehat{\mathbf{g}}$, $\widehat{\bsigma}$ and
$\widehat{\bsigma}_n$ we propose the following (W)ALS algorithm.
\begin{enumerate}
\item \emph{Initialization} Set the iteration counter $i = 1$ and initialize
  $\widehat\bsigma^{[0]}$ based on knowledge of $\bsigma_s$ and directional
  response of the sensors. For WALS, define the weight 
  $\mathbf{W}_c = \widehat\mathbf{R}^{-\frac{1}{2}}$.
  Initialize $\widehat{\mathcal{L}}^{[0]}$ based on knowledge of the nominal 
  position of the calibrator sources.
\item \emph{Estimate $\widehat{\mathbf{g}}^{[i]}$} by an eigenvalue
  decomposition of the matrix $\mathbf{M}$ with its entries obtained by
  averaging \eqref{eq:gratio} (resp.\ \eqref{eq:gratiopinv}) over all $k \neq
  i,j$ or by an eigenvalue decomposition of the matrix $\bC_2^\dagger \bC_1$
  as given by Eq.\ \eqref {eq:pinvC2C1} using $\mathbf{A}$ and
  $\widehat\bsigma^{[i-1]}$ as prior knowledge. Note that neither gain
  calibration approach does require knowledge of the sensor noise powers
  $\bsigma_n$ and that the latter approach is advisable if some elements may
  have a very low gain.
\item \emph{Estimate $\widehat\bsigma_n^{[i]}$} using either
  \eqref{eq:sigman1} or \eqref{eq:sigman2} 
  (resp.\ \eqref{eq:sigman1w} or \eqref{eq:sigman2w})
  applying available
  knowledge of $\widehat\mathbf{g}^{[i]}$, $\widehat\bsigma^{[i-1]}$ and
  $\mathbf{A}$.
\item \emph{Estimate $\widehat\bsigma^{[i]}$} using \eqref{eq:sigma}
    (resp.\ \eqref{eq:sigmaw}) and
  knowledge of $\widehat\mathbf{g}^{[i]}$, $\widehat\bsigma_n^{[i]}$ and
  $\mathbf{A}$.
\item\label{item:step5} If the DOAs are inaccurately known (scenarios 3 and
  4), \emph{estimate $\widehat{\mathcal{L}}^{[i]}$} using WSF as described in
  subsection \ref{subsec:WSF} using knowledge of $\widehat{\mathbf{g}}^{[i]}$
  and $\widehat\bsigma_n^{[i]}$ and initial estimate
  $\widehat{\mathcal{L}}^{[i-1]}$.
\item \emph{Check for convergence or stop criterion} 
  If $|\btheta^{[i-1]\dagger} \btheta^{[i]} - 1|
  < \delta$ or $i > i_{max}$, stop, otherwise increase $i$ by 1 and continue
  with step 2. 
\end{enumerate}
  The proposed criterion for convergence is based on a measure of
  the average relative error in all parameters, and will work
  even if the parameter values differ by orders of magnitude.

  The extension with step \ref{item:step5} is referred to as the Extended ALS
  (xALS) resp.\ xWALS algorithm.

  An algorithm that alternatingly optimizes for distinct groups of parameters
  can be proven to converge if the value of the cost function decreases in
  each iteration. In \cite{Viberg1991-1, Viberg1991-2} it is demonstrated that
  WSF minimizes the least squares cost function w.r.t.\ the parameterization
  of $\bA$, thus providing a partial solution to the least squares problem
  considered here. In step 2, $\widehat{\bg}$ is estimated using a method that
  only provides a near optimal solution to the least squares cost
  function. Its solution could, however, not be discerned from the true
  solution in Monte Carlo simulations, as demonstrated later in this
  paper. Optionally, step 2 could be augmented with one or two iterations of
  Eq.\ \eqref{eq:glin} to assure minimization of the least squares cost
  function. Alternatively, one could use the proposed estimate in the first
  iteration and use Eq.\ \eqref{eq:glin} in consecutive iterations in which a
  proper initial estimate is available from the previous iteration. The other
  parameters are estimated using well known standard solutions for least
  squares estimation problems. Therefore the value of the cost function is
  reduced in each step, thus ensuring convergence.

\subsection{Computational complexity}
\label{ssec:complexity}

\begin{table}
\renewcommand{\arraystretch}{1.3}
\begin{minipage}[b]{1.0\linewidth}
\centering
\begin{tabular}{llr}
\hline
\hline
ALS & $\widehat\mathbf{g}$ & $4\frac{1}{2}p^3 + pq^2 + o(\cdots)$\\
& $\widehat\bsigma$ & $2p^2q + 12pq^2 + 9q^3 + o(\cdots)$\\
& $\widehat\bsigma_n$ & 0\\
& $\widehat\sigma_n^2$ & 0\\
& \textbf{total} & $4\frac{1}{2}p^3 + 2p^2q + 13pq^2 + 9q^3 + o(\cdots)$\\
\hline
WALS & $\widehat\mathbf{g}$ & $5p^3 + pq^2 + o(\cdots)$\\
& $\widehat{\bg}$ (Eq.\ \eqref{eq:glin}) & $N_{iter} \left ( 2 \frac{2}{3} p^3
  + 4 p^2q + 2 q^2p + o(\cdots) \right )$\\
& $\widehat\bsigma$ & $\frac{2}{3}p^3+ 2p^2q + pq^2 + \frac{2}{3}q^3 +
o(\cdots)$\\
& $\widehat\bsigma_n$ & $\frac{5}{3}p^3 + o(\cdots)$\\
& $\widehat\sigma_n^2$ & $p^3 + o(\cdots)$\\
& \textbf{total} & $7\frac{1}{3}p^3 + 2p^2q + 2pq^2 + \frac{2}{3}q^3 +
o(\cdots)$
($\widehat\bsigma_n$)\\
& & $6\frac{2}{3}p^3 + 2p^2q + 2pq^2 + \frac{2}{3}q^3 + o(\cdots)$
($\widehat\sigma_n^2$)\\
\hline
\hline
\end{tabular}
\caption{Complexity of ALS and WALS iterations. \label{tab:complexity}}
\end{minipage}
\end{table}

Table \ref{tab:complexity} summarizes the numerical complexity of different
stages of the ALS and WALS algorithms per iteration expressed in the number of
complex multiplications. $N_{iter}$ is the number of iterations required for
convergence when using Eq.\ \eqref{eq:glin}. In this table, the notation
$o(\cdots)$ is used to denote all the neglected lower order terms. The
complexity of the omnidirectional gain estimates is dominated by the
eigenvalue decomposition on $\mathbf{M}$, which takes approximately $4p^3$
multiplications assuming use of the divide and conquer method
\cite{Wikipedia}. This step dominates the overall complexity as well, so it
may be worthwhile to estimate the number of iterations of the power method
\cite{Moon2000-1} required to obtain sufficient accuracy, especially if $p$ is
large.

The pseudo-inverse in Eq.\ \eqref{eq:sigma} can be implemented by a singular
value decomposition which is computationally more efficient than direct
computation of the Moore-Penrose inverse \cite{Moon2000-1}. Some intermediate
results, such as $\mathbf{G} \mathbf{A} \bSigma \mathbf{A}^H \mathbf{G}^H$,
are required more than once. In our calculation we assume that these terms are
computed only once and stored for future use. Under this assumption the number
of complex multiplications required to compute the noise power reduces to
zero. The number of array elements $p$ will generally be considerably larger
than the number of source signals $q$. Therefore the $q^3$-, $pq^2$- and
$p^2q$-terms may be considered negligible compared to the $p^3$-term in most
cases. These results suggest that the better statistical performance of the
WALS algorithm compared to the ALS algorithm as demonstrated by the simulation
results presented in the next section comes with only a minor increase in
complexity.

Both algorithm can be augmented with source position estimates using WSF. In
\cite{Viberg1991-2} it is demonstrated that WSF can be implemented with
$O(pq^2)$ complex operations per Gauss-Newton-type iteration of the proposed
modified variable projection algorithm once the result from the eigenvalue
decomposition is available. The cost of WSF is therefore dominated by the
eigenvalue decomposition on $\widehat\mathbf{R}$ requiring about $4p^3$
complex multiplications. This makes the computational cost of source location
estimation comparable to the cost of estimating the direction independent
gains.

\section{Simulation results}
\label{sec:simulations}

\begin{table}
\renewcommand{\arraystretch}{1.3}
\caption{Source powers and source locations used in the simulations}
\label{tab:srcmodel}
\begin{minipage}[b]{1.0\linewidth}
\centering
\begin{tabular}{l|rrr}
\hline
\hline
\# & $l$ & $m$ & $\sigma_q^2$\\
\hline
1 & 0.24651 & -0.71637 & 1.00000\\
2 & -0.34346 & 0.76883 & 0.88051\\
3 & -0.13125 & -0.31463 & 0.79079\\
4 & -0.29941 & -0.52339 & 0.74654\\
5 & 0.39290 & 0.58902 & 0.69781\\
\hline
\hline
\end{tabular}
\end{minipage}
\end{table}

The methods proposed in the previous section were tested using Monte Carlo
simulations and compared with the CRB. For these simulations a five-armed
array was defined, each arm being an eight-element one wavelength spaced
ULA. The first element of each arm formed an equally spaced circular array
with half wavelength spacing between the elements. The source model used in
the simulations is presented in Table \ref{tab:srcmodel}. This source model
was created using random number generators to demonstrate that the proposed
approach indeed works for arbitrary source models.

\begin{figure}
\centering
\includegraphics[width=8.5cm]{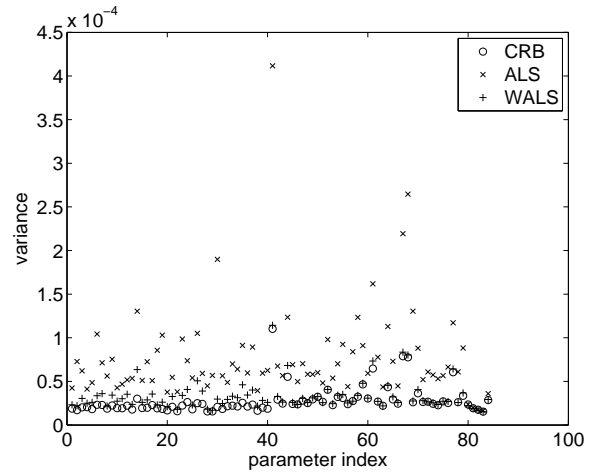}
\caption{Variance of the parameter vector $\btheta = [\bgamma^T, \phi_2,
  \cdots, \phi_{40},$ $\sigma_2^2, \cdots, \sigma_5^2, \sigma_n^2]^T$ estimates
  obtained in Monte Carlo simulations for the weighted and unweighted versions
  of the alternating least squares algorithm and compared these results with
  the CRB. \label{fig:MCsim8var}}
\end{figure}

In the first simulation the direction independent gains, the source powers and
the receiver noise powers were estimated assuming an array of equal elements,
i.e. the parameter vector was defined as $\btheta = [\bgamma^T, \phi_2,
\cdots, \phi_{40}, \sigma_2^2, \cdots \sigma_5^2, \sigma_n^2]^T$ corresponding
with the first scenario mentioned in section \ref{sec:datamodel}. Data was
generated assuming $\sigma_n^2 = 10$ and $N = 10^5$. This value for
$\sigma_n^2$ implies an instantaneous SNR on the strongest source of only 0.1
per array element, a typical situation for radio astronomers. Figure
\ref{fig:MCsim8var} shows the variance of the estimated parameters based on
1000 runs and compares these values with the CRB. As expected the WALS method
appears to be asymptotically statistically efficient while the ALS approach
does not attain the CRB and shows clear outliers.

\begin{figure}
\centering
\includegraphics[width=8.5cm]{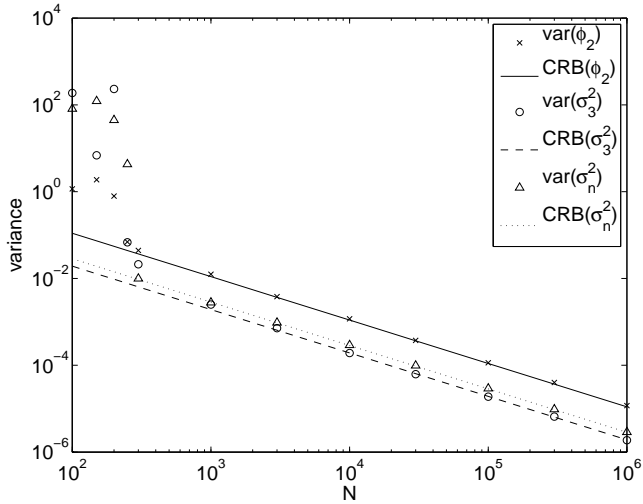}
\caption{Variances on $\phi_2$. $\sigma_3^2$ and $\sigma_n^2$ as function of
  the number of samples $N$, compared to the corresponding
  CRBs. \label{fig:MCsim_convergence}}
\end{figure}

\begin{figure}
\centering
\includegraphics[width=8.5cm]{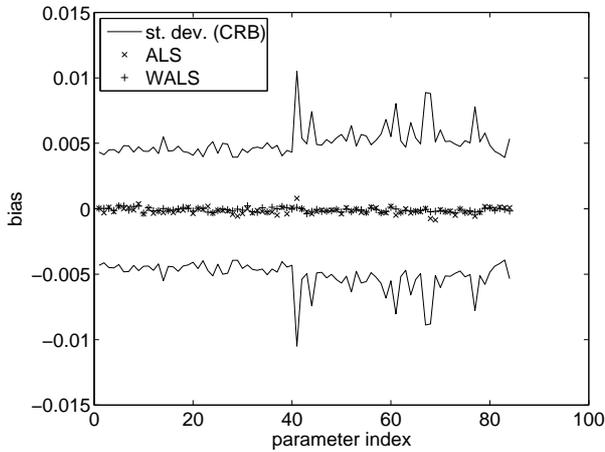}
\caption{Bias on the estimated parameters ($\btheta =
  [\bgamma^T, \phi_2, \cdots, \phi_{40},$ $\sigma_2^2, \cdots, \sigma_q^2,
  \sigma_n^2]^T$) for the weighted and unweighted version of the alternating
  least squares algorithm in the Monte Carlo simulations. The bias is compared
  with the standard deviation on the estimates derived from the
  CRB.\label{fig:MCsim8bias}}
\end{figure}

Figure \ref{fig:MCsim_convergence} shows the variance found on a number of
representative parameters as function of the number of samples and compares
this with the corresponding CRBs. This plot confirms the conjecture raised in
the previous paragraph that the WALS method is asymptotically statistically
efficient. Figure \ref{fig:MCsim8bias} shows the bias found in the Monte Carlo
simulations and compares this with the statistical error of 1$\sigma$ for a
single realization based on the CRB. This result indicates that both methods
are unbiased for all parameters.

\begin{figure}
\centering
\includegraphics[width=8.5cm]{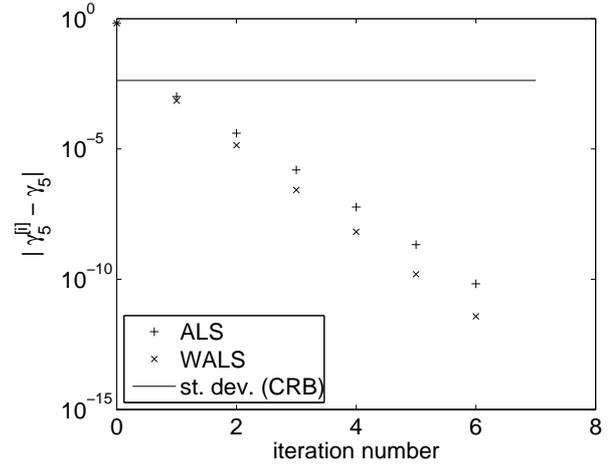}
\caption{The difference between the value of $\widehat\gamma_5$ at the end of
  each iteration and its final value is plotted versus the iteration number
  for ALS and WALS obtained in one of the Monte Carlo simulation runs. The
  behavior shown in this plot is representative for the results obtained for
  other parameters and from other runs. The standard deviation based on the
  CRB is also shown to demonstrate that one but preferably two iterations are
  sufficient to obtain a sufficiently accurate
  result. \label{fig:convergence}}
\end{figure}

Figure \ref{fig:convergence} shows the difference of the parameter value at
the end of each iteration and its final value obtained from one run of the
Monte Carlo simulation for a representative case, i.e.\ similar results were
found for other parameters and other runs. The standard deviation based on the
CRB is plotted as reference to facilitate the interpretation of the vertical
scale. The results indicate that only one or two iterations are needed to
reach the CRB in this scenario. With regard to other scenarios, this result
suggests exponential convergence, i.e. each iteration adds about one
significant digit to the parameter estimate, and it indicates that the WALS
method converges more rapidly than the ALS method. In these simulation the
stop criterion was $|\btheta^{[i-1]\dagger} \btheta^{[i]} - 1| < 10^{-10}$.

\begin{figure}
\centering
\includegraphics[width=8.5cm]{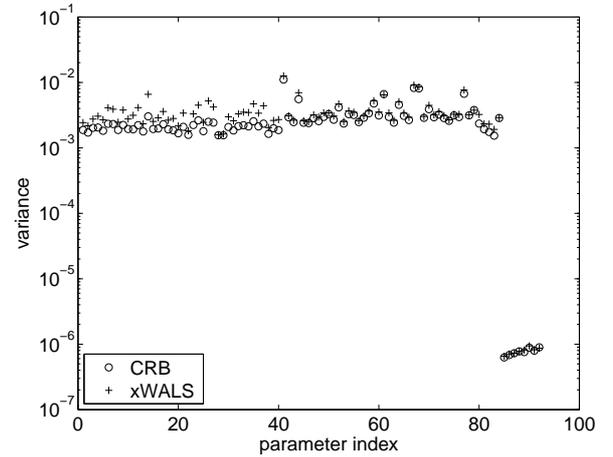}
\caption{Variance on the parameter vector $\btheta =
  [\bgamma^T, \phi_2, \cdots, \phi_{40},$ $\sigma_2^2, \cdots, \sigma_5^2,
  \sigma_n^2, l_2, \cdots, l_5, m_2 \cdots, m_5]^T$ estimates obtained in
  Monte Carlo simulations for the xWALS method, compared with
  the CRB. \label{fig:MCsim5var}}
\end{figure}

In the second simulation the direction independent gains, the apparent source
powers and source locations and the receiver noise powers were estimated
assuming an array of ideal elements. This corresponds to the third scenario
described in section \ref{sec:datamodel}. The parameter vector to be estimated
is thus $\btheta = [\bgamma^T, \phi_2, \cdots, \phi_{40}, \sigma_2^2, \cdots,
\sigma_5^2, \sigma_n^2, l_2, \cdots, l_5, m_2, \cdots, m_5]^T$. Data were
generated assuming $\sigma_n^2 = 10$ and $N = 1000$. This again implies an
instantaneous SNR of only 0.1 per array element but in this case the SNR per
receiver is only 3.2 for the strongest source after integration. Figure
\ref{fig:MCsim5var} shows the variance on the estimated parameters based on
1000 runs and compares these results with the CRB. In these simulations we
only used the xWALS algorithm, since the previous simulations demonstrated
that the ALS method gives inferior results compared to the WALS approach.
These results show that the variance on the parameter estimates are close to
the CRB.

\begin{figure}
\centering
\includegraphics[width=8.5cm]{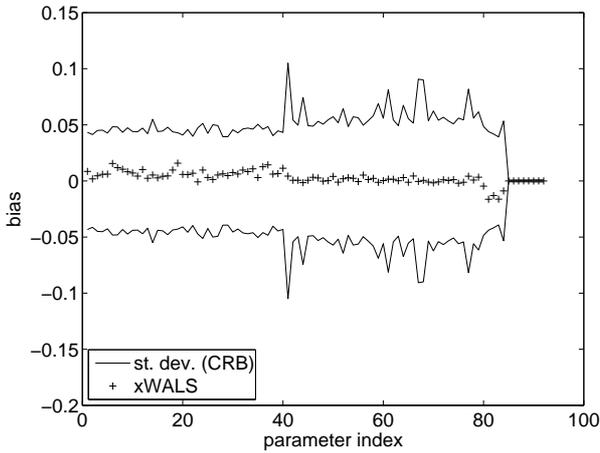}
\caption{Bias on the estimated parameters ($\btheta = [\bgamma^T, \phi_2,
  \cdots, \phi_{40},$ $\sigma_2^2, \cdots, \sigma_5^2, \sigma_n^2, l_2,
  \cdots, l_5, m_2 \cdots, m_5]^T$) for the xWALS algorithm obtained in Monte
  Carlo simulations. The results are compared with the standard deviation on
  the estimated derived from the CRB.\label{fig:MCsim5bias}}
\end{figure}

Figure \ref{fig:MCsim5bias} shows the bias on the estimated parameters found
in these Monte Carlo simulations. These results indicate that the estimates
are unbiased.

\begin{figure}
\centering
\includegraphics[width=8.5cm]{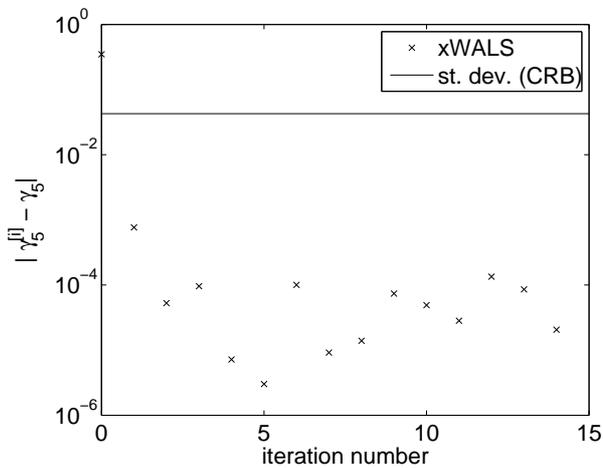}
\caption{The difference between the value of $\widehat\gamma_5$ at the end of
  each iteration and its final value is plotted versus the iteration number
  for one of the Monte Carlo simulation runs. The behavior shown in this plot
  is representative for the results obtained for other parameters and from
  other runs. The standard deviation based on the CRB is also shown to
  demonstrate that one but preferably two iterations are sufficient to obtain
  a sufficiently accurate result. \label{fig:convergence2}}
\end{figure}

The convergence of a representative parameter is shown in Fig.\
\ref{fig:convergence2}. Again one but preferably two iterations are sufficient
to get accurate results. However, in this scenario the stop criterion
$|\btheta^{[i-1]\dagger} \btheta^{[i]} - 1| < 10^{-10}$ was not reached and
the algorithm stops because the maximum number of iterations. which was set to
15, was reached. This indicates that the algorithm tries to interpret the
noise as real signal. In the first iteration first the omnidirectional gains,
apparent source powers and noise power are estimated. These values are then
used in the WSF to find the source locations. The source locations are then
used to update the sky model, leading to an update of $\widehat\bgamma$,
$\widehat\bsigma$ and $\widehat\bsigma_n$ and the cycle is complete. In the
first simulation, the SNR after integration defined as
$\sqrt{N}\sigma_i^2/\sigma_n^2$ was a factor 10 better than in this simulation
in which the SNR per array element per source after integration is only 2.2 to
3.2.

\section{Summary and conclusions}
\label{sec:conclusions}

In this paper we have developed a weighted alternating least squares (WALS)
algorithm to solve for direction independent complex gains, apparent source
powers and receiver noise powers simultaneously in a snapshot observation by a
sensor array. Our solution for finding the direction independent gains extends
and improves the available methods in the literature in the case of multiple
calibrator sources and provides robustness to small entries in the array
covariance matrix and small gain values (due to e.g.\ failing
elements). Although we have assumed independent sensor noise powers, our gain
estimation method is easily applied to cases in which some pairs of elements
experience correlated noise. We also found that unbiased estimation of the
receiver noise powers requires weighting with the best available array
covariance matrix model instead of weighting with the measured array
covariance matrix, which is common practice in signal processing.

The statistical performance was compared with an unweighted alternating least
squares (ALS) algorithm and the CRB and found to provide a statistically
efficient estimate thereby outperforming the ALS method. The computational
complexity of ALS and WALS scale with $4\frac{1}{2}p^3$ and about $7p^3$
respectively assuming that the number of calibrator sources is much smaller
than the number of array elements, i.e.\ the WALS method comes with only a
minor increase in computational burden. Simulations indicate that only one or
two iterations are sufficient to reach the CRB and that every iteration adds
about one significant digit, the WALS method converging slightly faster than
the ALS method.

We extended these methods with source location estimation using weighted
subspace fitting. Simulations indicate that this extension to the WALS
algorithm provides a statistically efficient simultaneous estimate of
omnidirectional gains, apparent source powers, source locations and receiver
noise powers. Again, one to two iterations proved to be sufficient to reach
the CRB. However, the convergence of the parameter is blocked at some level
lower than the CRB by the noise in the measurement, at which point the
algorithm keeps jumping from one local optimum to another.

\section{Acknowledgments}

This work has been carried out in close collaboration with the
LOFAR-team. This helped to develop our ideas up to the point at which our
method could be implemented to calibrate the LOFAR stations. Especially the
interaction with Albert-Jan Boonstra and Jaap Bregman at ASTRON on LOFAR
station calibration is gratefully acknowledged. The authors would also like to
thank the reviewers whose comments helped to improve the original manuscript.

\bibliography{refs}

\begin{biography}{Stefan J. Wijnholds}
  (S'2006) was born in The Netherlands in 1978. He received M.Sc.\ degrees in
  Astronomy and Applied Physics (both cum laude) from the University of
  Groningen in 2003. After his graduation he joined the R\&D department of
  ASTRON, the Netherlands Institute for Radio Astronomy, in Dwingeloo, The
  Netherlands, where he works with the system design and integration group on
  the development of the next generation of radio telescopes. Since 2006 he is
  also with the Delft University of Technology, Delft, The Netherlands, where
  he is pursuing a Ph.D.\ degree. His research interests lie in the area of
  array signal processing, specifically calibration and imaging.
\end{biography}

\begin{biography}{Alle-Jan van der Veen}
  (F'2005) was born in The Netherlands in 1966. He received the Ph.D.\ degree
  (cum laude) from TU Delft in 1993. Throughout 1994, he was a postdoctoral
  scholar at Stanford University.  At present, he is a Full Professor in
  Signal Processing at TU Delft.

  He is the recipient of a 1994 and a 1997 IEEE Signal Processing Society
  (SPS) Young Author paper award, and was an Associate Editor for IEEE Tr.\
  Signal Processing (1998--2001), chairman of IEEE SPS Signal Processing for
  Communications Technical Committee (2002-2004), and Editor-in-Chief of IEEE
  Signal Processing Letters (2002-2005).  He currently is Editor-in-Chief of
  IEEE Transactions on Signal Processing, and member-at-large of the Board of
  Governors of IEEE SPS.

  His research interests are in the general area of system theory applied to
  signal processing, and in particular algebraic methods for array signal
  processing, with applications to wireless communications and radio
  astronomy.
\end{biography}

\end{document}